\documentclass[prl,showpacs,twocolumn,aps]{revtex4}
\usepackage{rotating}
\usepackage{epsfig}
\newcommand{\be}{\begin{eqnarray}}

\newcommand{\ee}{\end{eqnarray}}
\newcommand{\bea}{\begin{eqnarray}}
\newcommand{\eea}{\end{eqnarray}}

\begin{document}
\title{Electron-phonon vs. electron-impurity interactions with small electron bandwidths}
\author{F. Do\u gan and F. Marsiglio}

\affiliation{Department of Physics, University of Alberta, Edmonton,
Alberta, Canada T6G 2J1}

\date{\today}
\begin{abstract}
It is common practice to try to understand electron interactions in
metals by defining a hierarchy of energy scales. Very often, the
Fermi energy is considered the largest, so much so that frequently
bandwidths are approximated as infinite. The reasoning is that
attention should properly be focused on energy levels near the Fermi
level, and details of the bands well away from the Fermi level are
unimportant. However, a finite bandwidth can play an important role for
low frequency properties: following a number of recent papers, we examine
electron-impurity and electron-phonon interactions in bands with
finite widths. In particular, we examine the behaviour of the
electron self energy, spectral function, density of states, and
dispersion, when the phonon spectral function is treated
realistically as a broad Lorentzian function. With this phonon
spectrum, impurity scattering has a significant non-linear effect.
\end{abstract}
\maketitle
\date{\today}

\section{I. INTRODUCTION}
\label{sec:int}

In the last few years, a number of authors have investigated the
effect of a finite Fermi energy on the properties of an interacting
electron system
\cite{alexandrov87,marsiglio92,dogan03,cappelluti03,knigavko04,knigavko05b,knigavko05c}.
In the past, with the understanding that only properties near the
Fermi level were important, the Fermi energy was assumed to be
infinite, to facilitate integrations over electronic energies
\cite{marsiglio03}. This approximation was believed to have
negligible impact on properties concerning states near the Fermi
level. Alexandrov et al. \cite{alexandrov87} first noted that
accounting for a finite bandwidth leads to a significant
modification of the electron density of states (EDOS) near the Fermi
level; they interpreted this as a ``polaronic collapse" of the electron band.
From our point of view the important observation is that the EDOS
actually experiences a significant alteration with electron-phonon
coupling. In this paper, following the work presented in Refs.
\onlinecite{dogan03,cappelluti03}, we wish to investigate further
the impact of a finite density of states when both impurity
scattering and electron-phonon coupling are present. Specifically we
will examine the impact of a different bare EDOS (a Lorentzian
model), and the impact of a more realistic broad phonon spectrum (previous
work focussed on an Einstein spectrum).

The paper is organized as follows: in the following section we
briefly review the formalism to calculate the electron self energy.
The formulation we use allows for several properties to be
obtained explicitly and analytically.
Section III deals briefly with only electron-impurity
scattering. Significant changes occur at a very basic level because
the bandwidth is finite. For example, the self energy acquires a
real part \cite{cappelluti03}. Section IV shows a number of results for combined
electron-phonon and electron-impurity scattering, where both a
delta-function and broadened Lorentzian form is used for the phonon
spectrum. We also make a direct comparison with the results of Ref.
\cite{cappelluti03} to show the model dependence of the analysis.
We close with a Summary of our results.

\section{II. THE FORMALISM}
\label{sec:for}

The self-consistent Migdal approximation has been explained before in the cited references, so the reader is referred to them.
For an arbitrary electron-phonon spectral function, $\alpha^2F(\nu)$, the resulting equations for the self energy,
$\Sigma(\omega+i\delta)$, are as follows, at zero temperature: \bea {\rm Re} \Sigma(\omega + i\delta) = {1 \over 2 \tau} \int
d\epsilon \frac{N_\circ(\epsilon)}{ N_\circ(0)} {\rm ReG}(\epsilon,\omega + i\delta)+
\nonumber\\
\nonumber \\
 \int_0^\infty \, d\nu \, \alpha^2F(\nu) \biggl[ ({(N(\omega^\prime) - N(\omega-\nu) \over N_\circ(0)}\bigr) {2\omega
\over \omega^2 - (\nu + \omega^\prime)^2} + \nonumber \\
{N(\omega-\nu) \over N_\circ(0)} \, {\rm ln}|{\omega - \nu \over \omega + \nu}| \biggr] \label{resig} \eea and \be {\rm Im}
\Sigma(\omega+i\delta) = - {1 \over 2 \tau} {N(\omega)\over N_\circ(0)}
 -\pi \int_0^\infty \, d\nu \, \alpha^2F(\nu) \nonumber \\\biggl[
{N(\omega-\nu) \over N_\circ(0)} \theta(\omega-\nu) + {N(\omega+\nu) \over N_\circ(0)} (1-\theta(\omega + \nu))\biggr].
\label{imsig} \ee Here, the electron Green function, $G(\epsilon,\omega+i\delta)$, is given by the Dyson equation: \be
G(\epsilon,\omega+i\delta) = 1/[\omega + i\delta - \epsilon - \Sigma(\omega+i\delta)], \label{dyson} \ee where the momentum
dependence is summarized by the energy $\epsilon$. The parameter $1/\tau$ characterizes the impurity scattering strength, and has
units of a scattering rate. The bare electron density of states (EDOS) is given by $N_\circ(\epsilon)$, while the renormalized
density of states (RDOS) is given by $N(\omega)$. The former function in principle comes from a band structure calculation; we
will simply model it either as a constant with finite bandwidth or a Lorentzian \cite{dogan03}. The RDOS is a product of our
calculation, and is given by \be N(\omega) = \int_{-\infty}^{\infty} d\epsilon \ N_\circ(\epsilon) A(\epsilon,\omega),
\label{ren_dense} \ee where $A(\epsilon,\omega) \equiv -{1 \over \pi} {\rm Im} G(\epsilon,\omega + i \delta)$ is the electron
spectral function. For the two simple models for the bare EDOS, $N(\omega)$ can be determined analytically in terms of the
electron self energy \cite{dogan03}. Thus, these equations constitute a set of self-consistent equations for the electron self
energy, from which all single electron properties can be calculated. As already mentioned, for the electron-impurity scattering,
the process is characterized by a single number, $1/\tau$. On the other hand, the electron-phonon interaction is specified by an
entire function, $\alpha^2F(\nu)$, which we will sometimes take to be an Einstein spectrum, in which case two parameters are
required, the frequency, and the ``strength", as measured by, say, the mass enhancement parameter, $\lambda \equiv 2\int_0^\infty
d\nu \ {\alpha^2F(\nu) \over \nu}$.  To specifically examine the impact of broadened spectra, we will utilize the truncated
Lorentzian shape defined in Ref. \onlinecite{dogan03}.

Note that an alternative formulation first determines the relevant
Green function and self energy on the imaginary axis; then a set of
self-consistent equations can be utilized to analytically continue
the result to just above the real axis \cite{marsiglio88}. This
method was used in
Ref.~(\onlinecite{cappelluti03},\onlinecite{knigavko05c}), and is
particularly useful at finite temperature.

\section{III. ELECTRON-IMPURITY SCATTERING}
\label{sec:el-imp}

Eqs. (\ref{resig}-\ref{ren_dense}) require iterative self-consistent solutions. Analytic results
are not possible, except in some limits. For example, with impurity scattering only, and with
a bare EDOS given by a Lorentzian with half-width $D/2 \equiv \epsilon_F$, the self energy
is given by
\be
\Sigma(\omega + i \delta) = {\omega + i\epsilon_F \over 2} \biggl(
1 - \sqrt{1 - {2 \epsilon_F/\tau \over (\omega + i\epsilon_F)^2}}
\biggr).
\label{sig_imp}
\ee
\begin{figure}[tp]
\begin{center}
\begin{turn}{-90}
\epsfig{figure=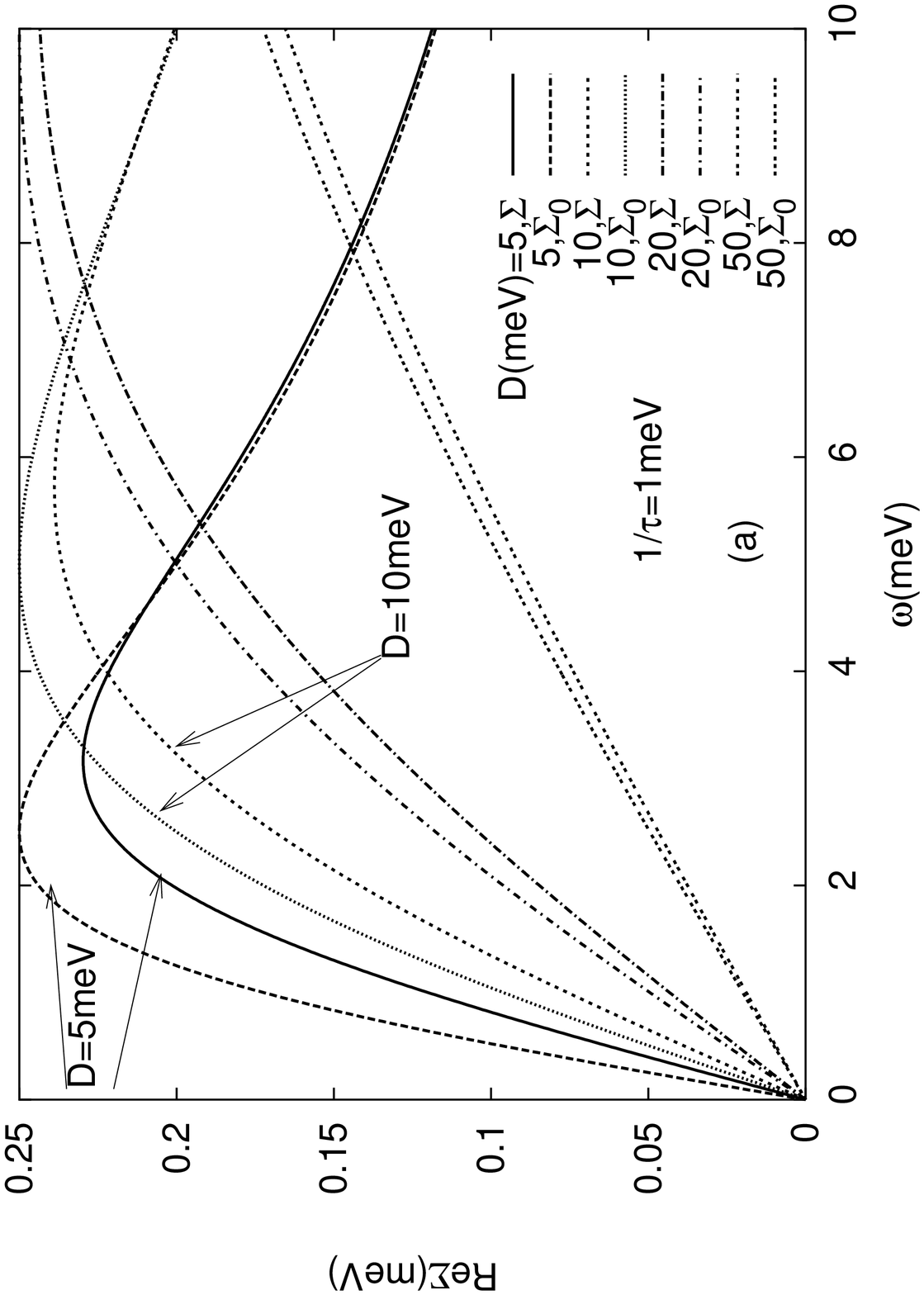,height=3.5in} \epsfig{figure=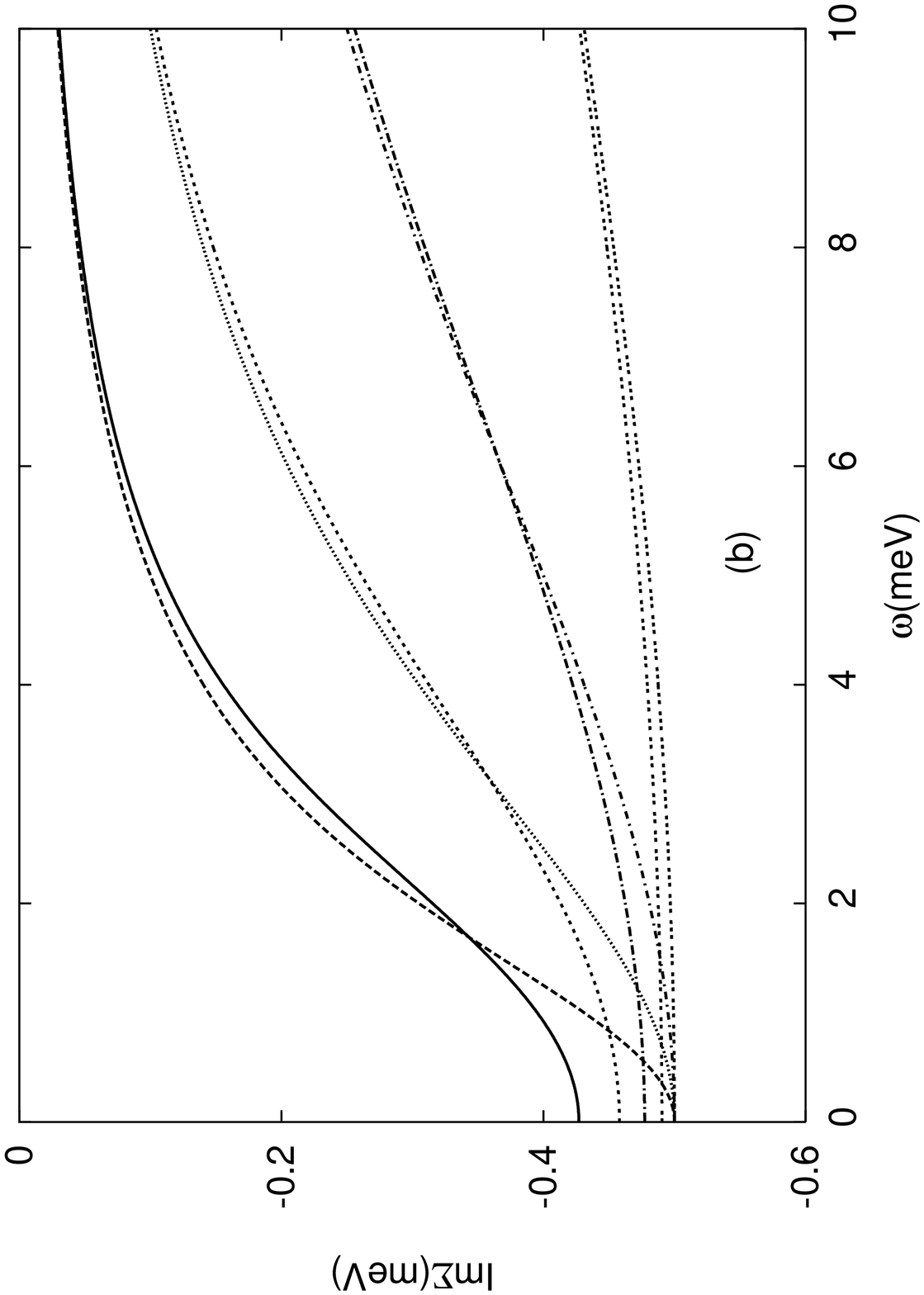,height=3.5in}
\end{turn}
\caption{Real (a) and imaginary (b) parts of the electron self-energy, with an electron-impurity scattering strength $1/\tau = 1$
meV. The results for different bandwidths are shown, for both the fully self consistent result ($\Sigma$) and for the non-self
consistent result ($\Sigma_0$). The legend for both parts is given in (a). For infinite bandwidth the real part is zero, and the
imaginary part is a constant at $-1/(2\tau)$ (not shown). Note that full self consistency alters the scattering rate at the Fermi
level.}
\end{center}
\end{figure}
For an infinite bandwidth the impurity self energy is pure imaginary, $\Sigma(\omega + i \delta)
= -i{1 \over 2\tau}$; finally an approximate result is obtained by using the first term in
Eqs. (\ref{resig},\ref{imsig}) with the {\it bare} Green function on the right hand side.
These results
are illustrated in Fig.~1. Similar results are obtained for a constant bare EDOS with finite
bandwidth \cite{knigavko05b}. For the imaginary part of the self energy (which Eq. \ref{imsig}
indicates is proportional to the density of states when only impurity scattering is present)
the main effect of the finite bandwidth is to reduce the self energy to zero beyond energies
corresponding to the electronic states. The real part (Fig.~1a) experiences the biggest change,
since it is zero at all frequencies if the limit of infinite bandwidth is taken first.
Fig.~1b illustrates that self-consistency impacts the low frequency behaviour in particular.
Note that while the intercept of the imaginary part of the self energy retains the physical
interpretation of a scattering rate, the slope of the real part is no longer related to an
effective mass. This will be pertinent when both electron-impurity and electron-phonon
scattering are present, as already pointed out in Ref. \onlinecite{cappelluti03}.

\begin{figure}[tp]
\begin{center}
\begin{turn}{-90}
\epsfig{figure=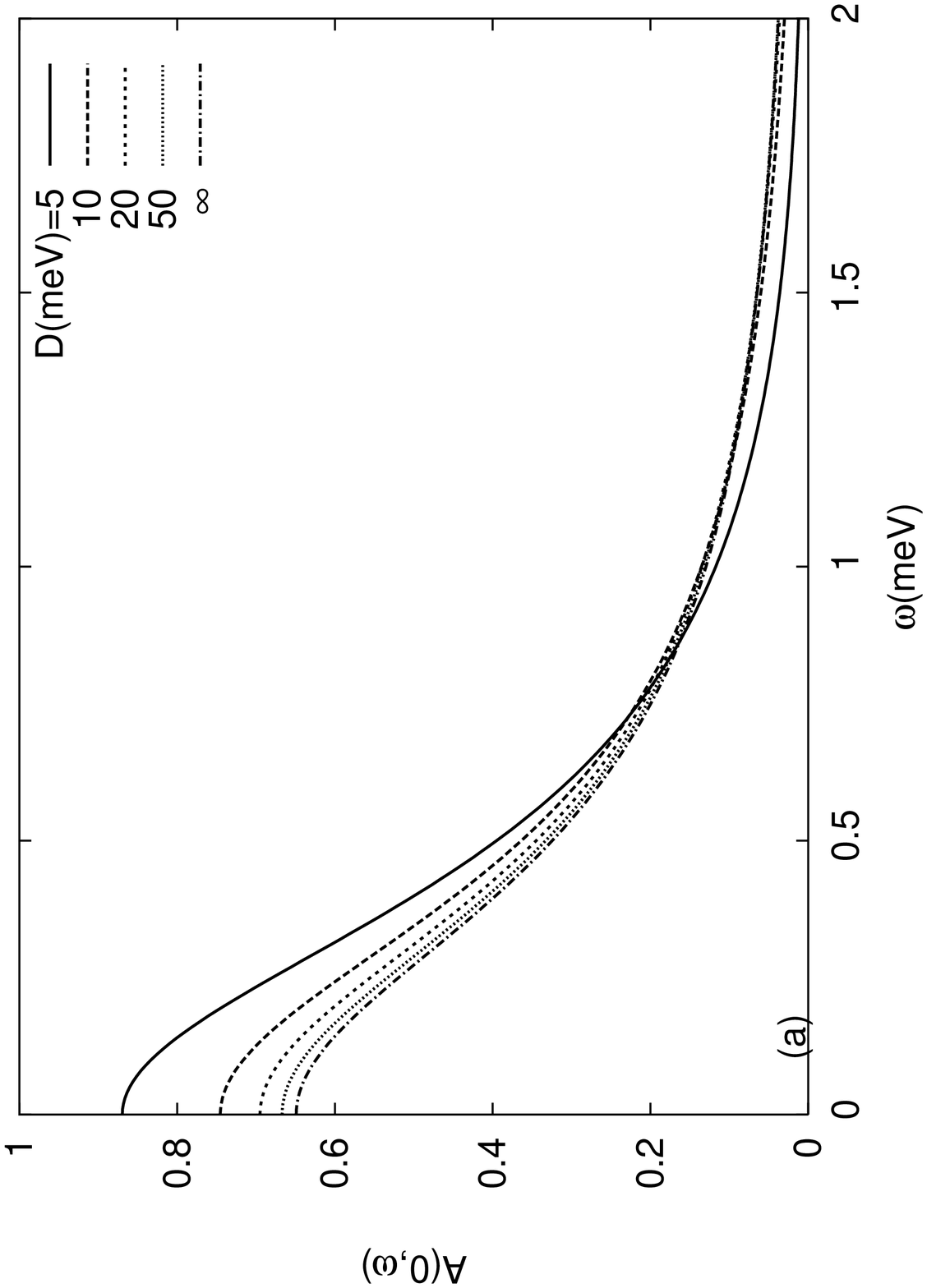,height=3.5in} \epsfig{figure=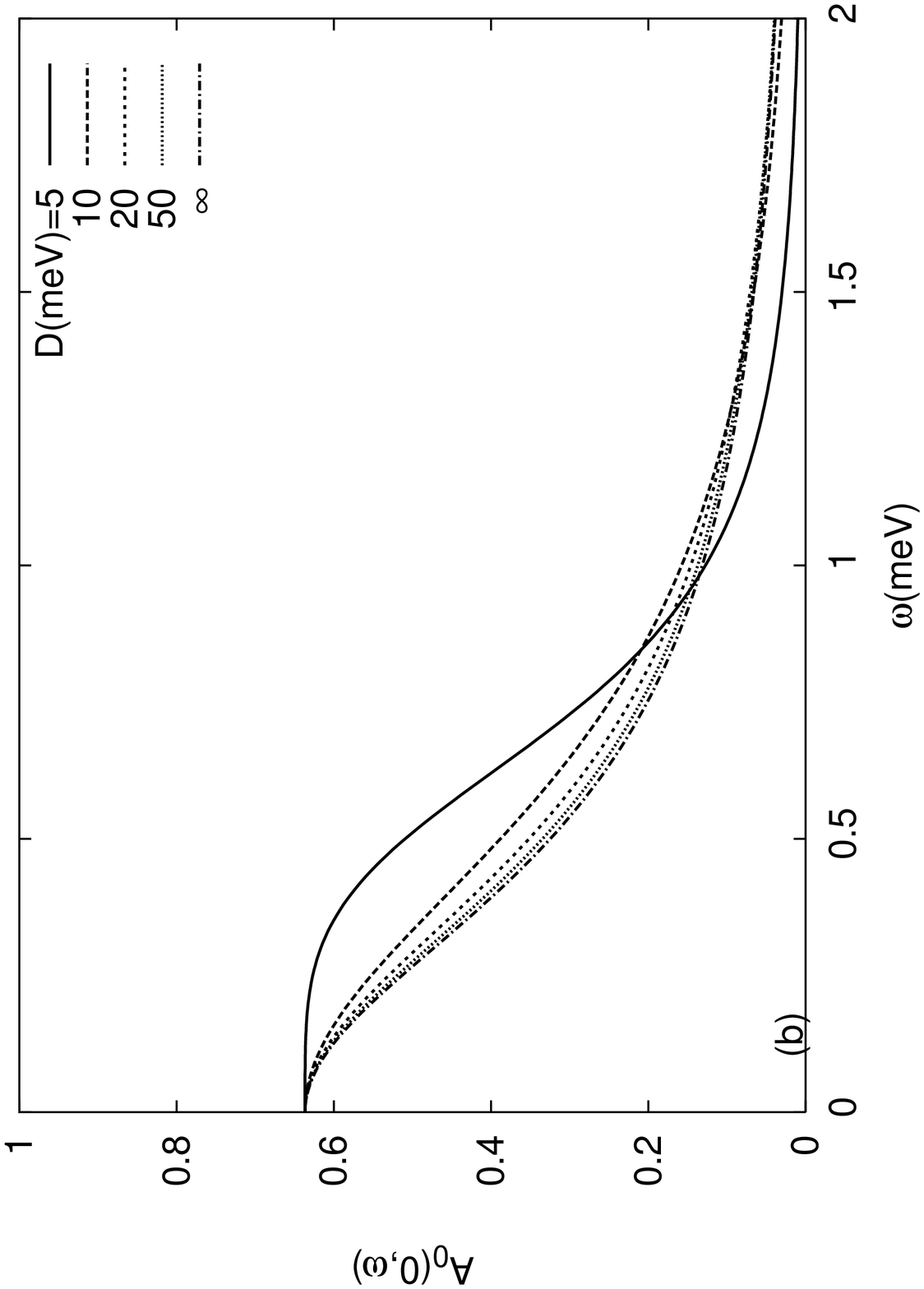,height=3.5in}
\end{turn}
\caption{The full self consistent (a) and non-self consistent (b) electron spectral function for various values of the bandwidth,
including the infinite bandwidth case. This latter result is of Lorentzian form, whereas all others deviate from this simple
form. Note that the non-self consistent results (b) can be very non-Lorentzian for small bandwidths.}
\end{center}
\end{figure}
The renormalized EDOS is already shown (to within a minus sign and a
constant factor --- see Eq. (\ref{imsig})) in Fig.~1b. Note the
renormalization at the Fermi level for the fully self-consistent
results. From Eq. \ref{ren_dense}, we know that the renormalization
of the EDOS is a convolution over the spectral function. Fig.~2
shows the spectral function at the Fermi level. As expected, the
fully self-consistent calculation (Fig.~2a) reflects the
renormalization whereas the non-self-consistent one does not
(Fig.~2b). The latter shows that the spectral shape can become very
non-Lorentzian when the Fermi energy becomes quite small. The former
shows that it can be quite difficult to unravel the impurity
scattering strength from the details of the electronic band
structure (and here we are simply varying the bandwidth). As we
shall see, the contributions from electron-phonon scattering
generate even more ambiguity.

\section{IV. ELECTRON-PHONON PLUS ELECTRON-IMPURITY SCATTERING}
\label{sec:imp-pho}

\section{(a) Einstein phonon}
\label{einstein}

\begin{figure}[tp]
\begin{center}
\begin{turn}{-90}
\epsfig{figure=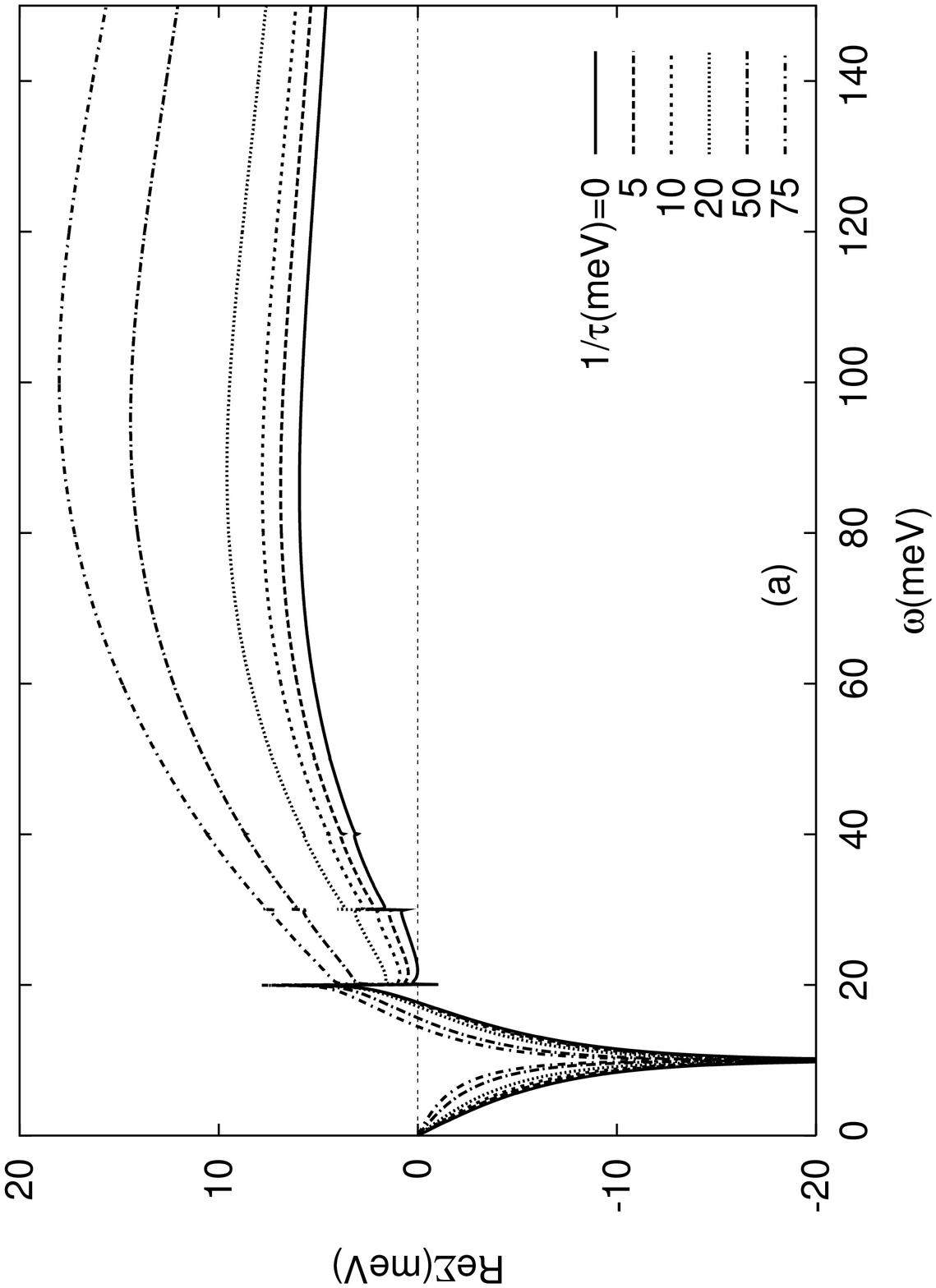,height=3.5in} \epsfig{figure=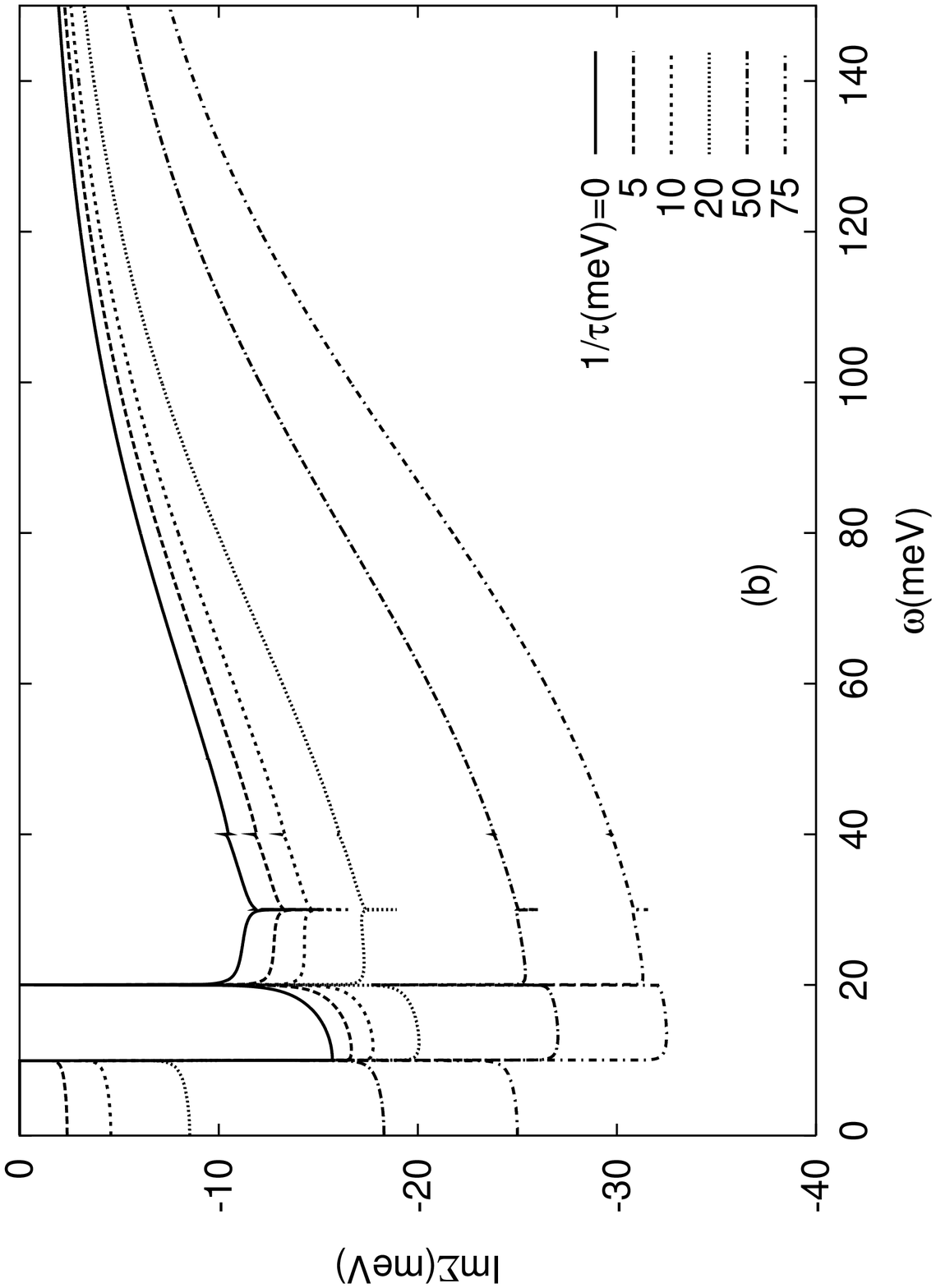,height=3.5in}
\end{turn}
\caption{Real (a) and imaginary (b) parts of the electron self-energy, for a moderate electron-phonon coupling strength, $\lambda
= 1$. The phonon spectrum is a delta function at frequency $\nu_E = 10$ meV. Results are shown for one bandwidth, $D = 10\nu_E$,
and for various electron-impurity scattering strengths, as shown. Note that the logarithmic singularity in (a) remains for all
impurity scattering strengths. Overall, the "sharp" features due to the Einstein oscillator remain throughout, independent of the
impurity scattering rate.}
\end{center}
\end{figure}
Results with just the electron-phonon interaction have already been
presented \cite{dogan03}, so we will focus on the two effects in
combination. Figs.~3a and 3b show the real and imaginary parts of
the self energy as the electron-impurity scattering strength is
varied. For the electron-phonon scattering, we have used an Einstein
spectrum with frequency $\nu_E = 10$ meV, and strength $\lambda =
1$. Our results are similar to those of Ref.
\onlinecite{cappelluti03} except that our real part has much sharper
features than theirs, presumably because ours is at $T=0$ (theirs is
at a small non-zero temperature). It is also evident from our method
of solution (see Eqs. (\ref{resig},\ref{imsig})) that a logarithmic
singularity at $\omega = \nu_E$ persists no matter what strength of
impurity scattering is present. Thus, somewhat counterintuitively,
impurity scattering does not smear out the singularity at $\omega =
\nu_E$. As will be evident below, the singularity at $\omega =
\nu_E$ results in a collapse of the renormalized EDOS at that
frequency. The phonon contribution to the imaginary part of the
self-energy is zero until $\omega = \nu_E$, and the impurity
contribution is also zero at $\omega = \nu_E$, due to the collapse
of the electron band at this frequency. This combination ensures
that the overall imaginary part of the self energy drops to zero at
this frequency, as Fig.~3b shows. However, at $\omega = 2\nu_E$, the
result is in general non-zero: while the electron-phonon
contribution is zero, since it depends on the EDOS with frequency
shifted by $\nu_E$, the impurity contribution is not zero since the
renormalized EDOS at that frequency is non-zero. Therefore the total
function is non-zero for $\omega = 2\nu_E$.

\begin{figure}[tp]
\begin{center}
\begin{turn}{-90}
\epsfig{figure=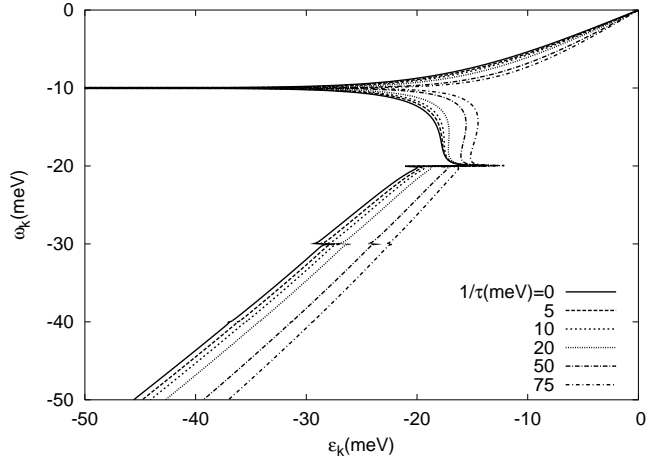,height=3.5in}
\end{turn}
\caption{The dispersion corresponding to the parameters of Fig.~3. As mentioned with respect to Fig.~3, sharp features remain in
spite of the impurity scattering.}
\end{center}
\end{figure}
As Cappelluti and Pietronero pointed out already in Ref.~(\onlinecite{cappelluti03}), the slope of the real part of the self
energy at the Fermi energy ($\omega = 0$) changes systematically with impurity scattering. This makes inference of the
electron-phonon coupling strength from the self energy a very difficult problem. Moreover, the ``kink" observed in photoemission
spectroscopy \cite{lanzara01} will be affected. However, for an Einstein spectrum, the ``kink" actually becomes more pronounced
with increasing impurity scattering, as shown in Fig. 4. This is because the logarithmic singularity remains in the real part of
the self energy, but impurity scattering makes this singularity sharper, as Fig.~4 illustrates. As shown below, with a more
realistic electron phonon coupling spectrum the conclusion of Ref. \onlinecite{cappelluti03} applies, namely that the ``kink"
becomes smeared with increasing electron-impurity scattering.

In Fig.~5a and 5b we show the spectral function at the Fermi energy and the renormalized
EDOS, respectively, as a function of frequency. Note the significant reduction near $\omega
=0$ with increased impurity scattering in both properties, with the compensating weight
gain occurring at high frequencies.

\begin{figure}[tp]
\begin{center}
\begin{turn}{-90}
\epsfig{figure=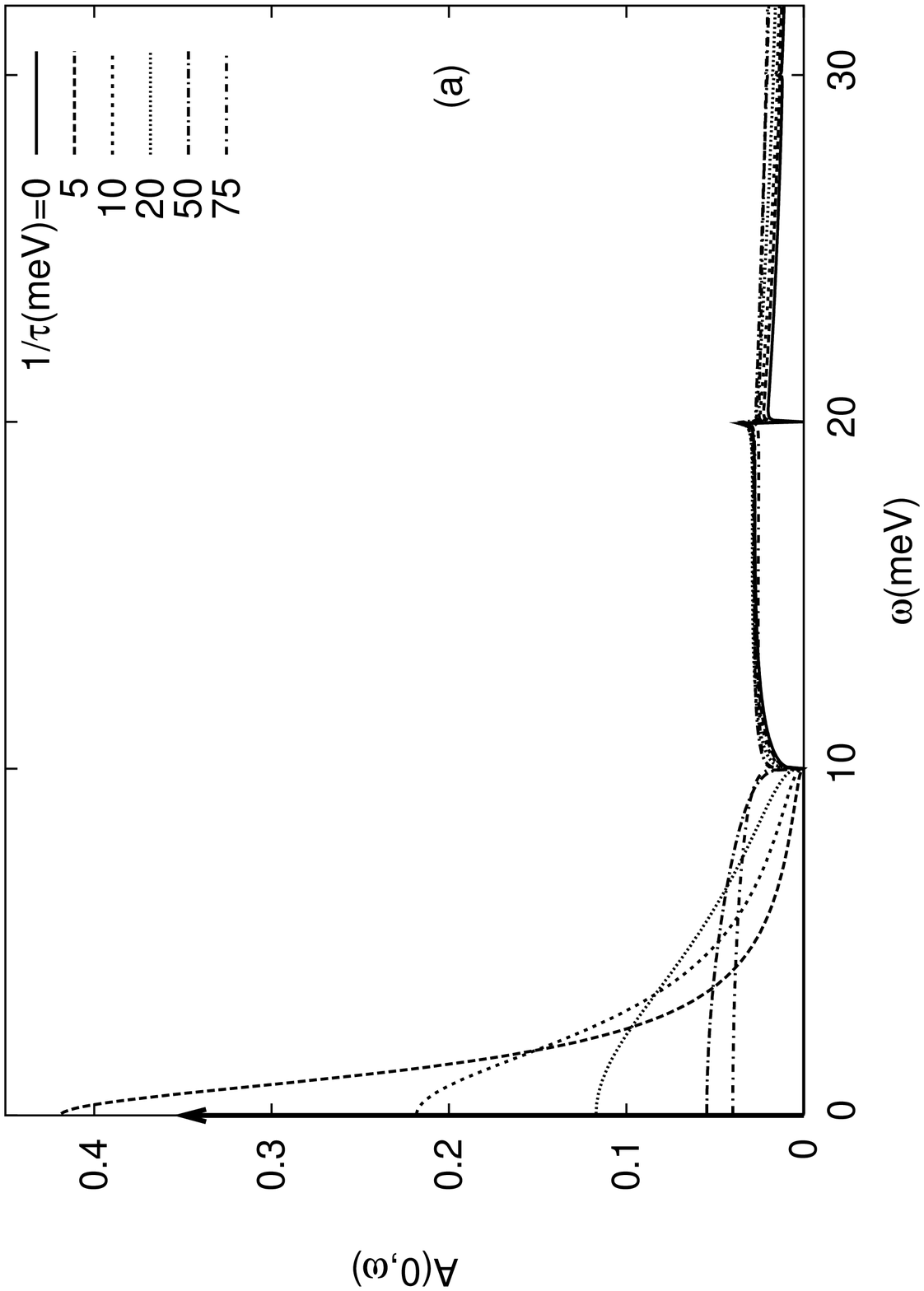,height=3.5in} \epsfig{figure=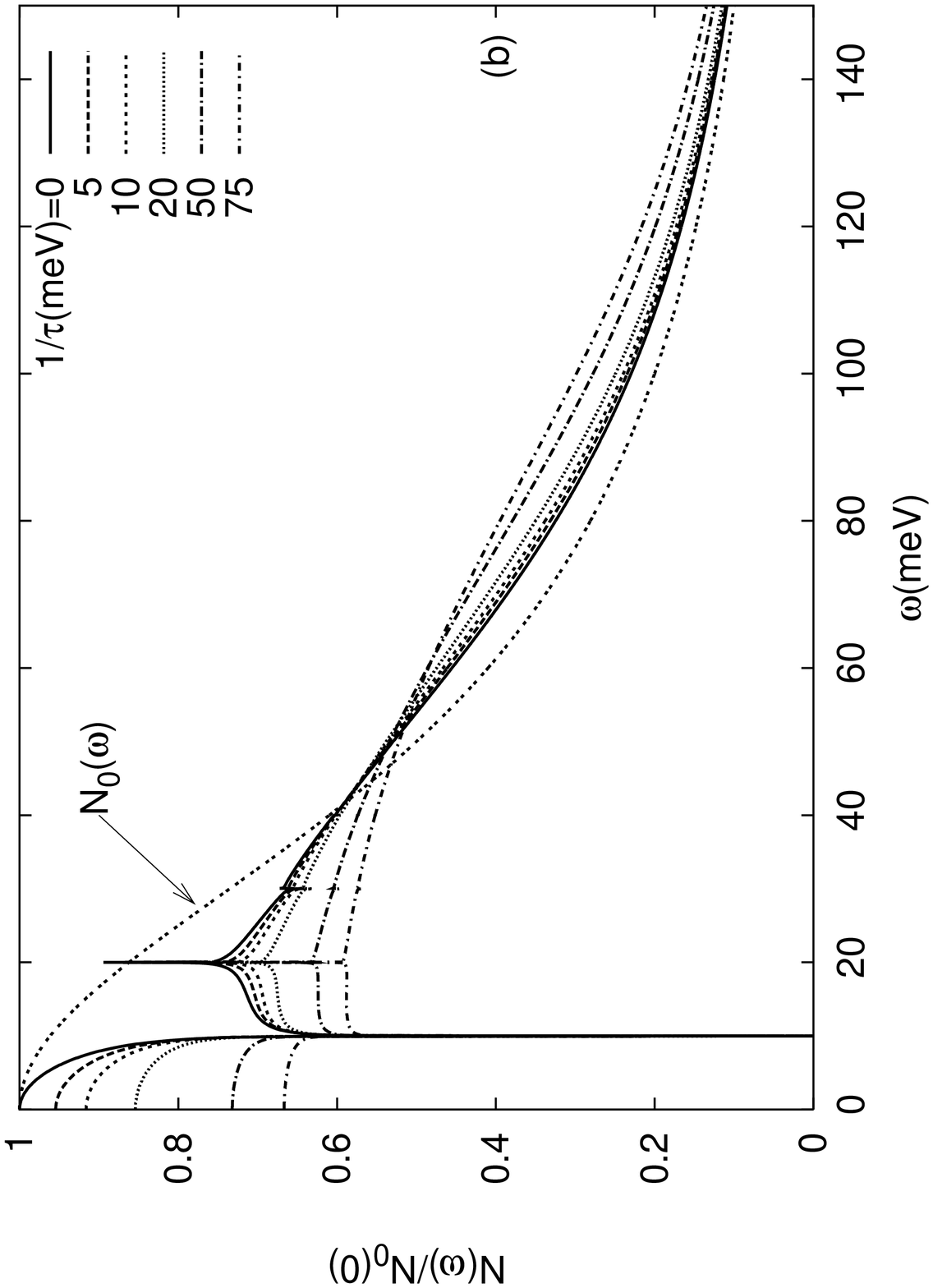,height=3.5in}
\end{turn}
\caption{The spectral function at the Fermi level (a) and the renormalized density of states for the parameters of Fig.~3, for
various degrees of impurity scattering. Note that in (b) the "collapse" of the band evident at $\omega = \nu_E$ remains for all
strengths of impurity scattering. Spectral weight is pushed to higher energies as the impurity scattering increases. In (a) when
no impurity scattering is present (solid curve) the spectral function is given by a delta function at the origin (indicated by
the vertical line with the arrow), followed by an incoherent piece that starts at $\omega = \nu_E = 10$ meV.}
\end{center}
\end{figure}
\section{(b) Lorentzian phonon spectrum}
\label{lorentzian}

\begin{figure}[tp]
\begin{center}
\begin{turn}{-90}
\epsfig{figure=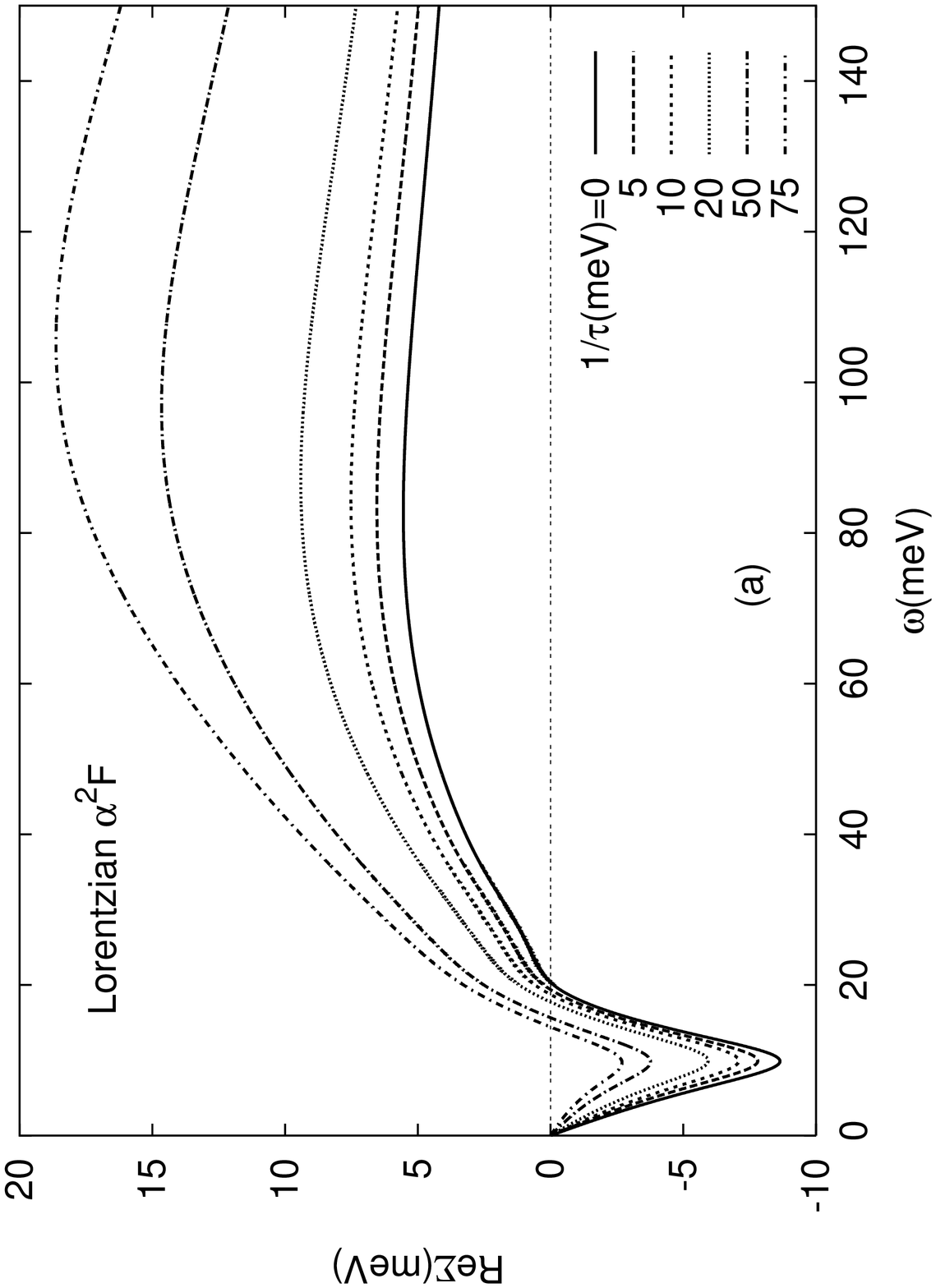,height=3.5in} \epsfig{figure=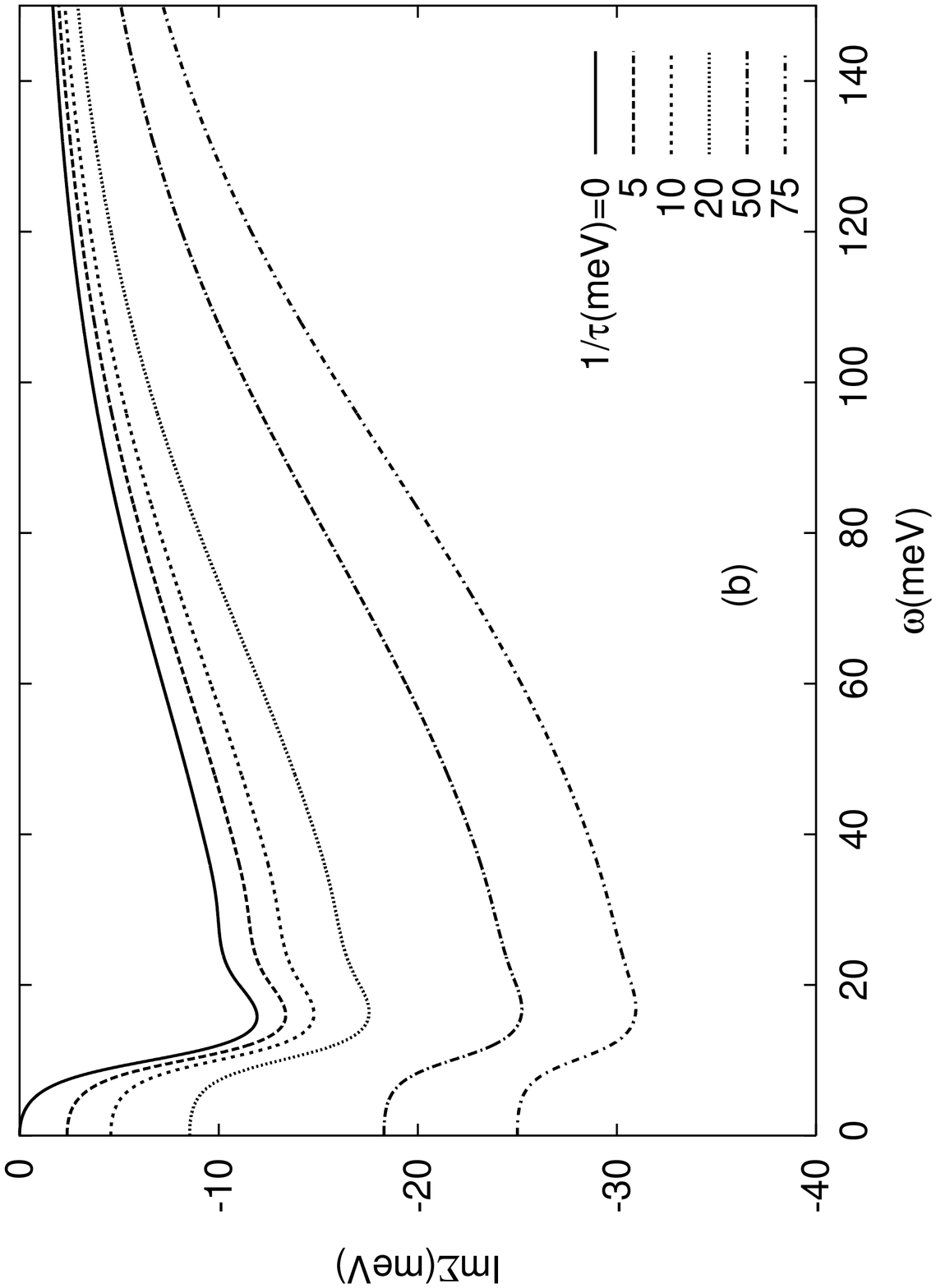,height=3.5in}
\end{turn}
\caption{Real (a) and imaginary (b) parts of the electron self energy, calculated using an interaction with phonons described by
a Lorentzian spectrum centered at $\omega = 10$ meV with $\lambda = 1$. The Lorentzian describing the phonons has broadening
parameter, $\delta = 3$ meV. Note that overall trends are the same as for the Einstein phonon (Fig.~3) except everything is
smooth. In the real part the degree of the dip (near $\omega \approx \nu_E = 10$ meV) is strongly altered by the impurity
scattering (in the Einstein phonon case it remains singular).}
\end{center}
\end{figure}
Many of these results contain anomalously sharp features because of
the use of an Einstein model for the phonon spectral function. It
was already shown in Ref. \onlinecite{dogan03} that these features
are considerably smoothened when a more realistic spectrum is used.
Here we use the broadened Lorentzian spectrum of Ref.
\onlinecite{dogan03} with $\delta = 3$ meV. Results are shown in
Fig. 6 for (a) the real and (b) imaginary parts of the self energy.
Clearly the singularity in the real part is absent even without
impurity scattering; in contrast to the Einstein mode case, however,
impurity scattering alters the nature of the dip near $\nu_E$
considerably. The rest of the frequency dependence for both real and
imaginary parts is also affected considerably, as was the case with
the Einstein spectrum. In particular, the high frequency portion of
the real part remains enhanced due to impurity scattering, and, as
remarked in Ref. \onlinecite{cappelluti03}, this will affect the
interpretation of the high frequency extrapolation of the dispersion
relation as measured in photoemission \cite{lanzara01}. Of special
note here, however, is the definite reduction of the magnitude of
the slope of the real part of the self energy near $\omega = 0$. The
way this would appear for the dispersion relation is shown in
Fig.~7. Note the ambiguity which impurity scattering causes for the
extraction of an electron-phonon coupling strength from such a
measurement \cite{cappelluti03}.
\begin{figure}[tp]
\begin{center}
\begin{turn}{-90}
\epsfig{figure=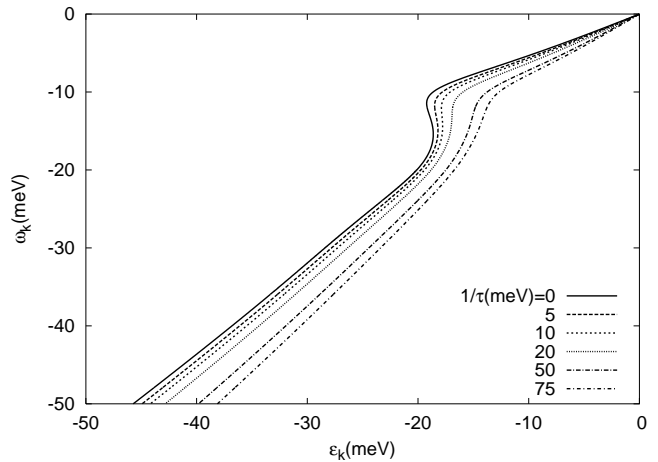,height=3.5in}
\end{turn}
\caption{The dispersion corresponding to the parameters of Fig.~6. The severity of the "kink" at $\omega = 10$ meV is
significantly moderated by impurity scattering.}
\end{center}
\end{figure}

\begin{figure}[tp]
\begin{center}
\begin{turn}{-90}
\epsfig{figure=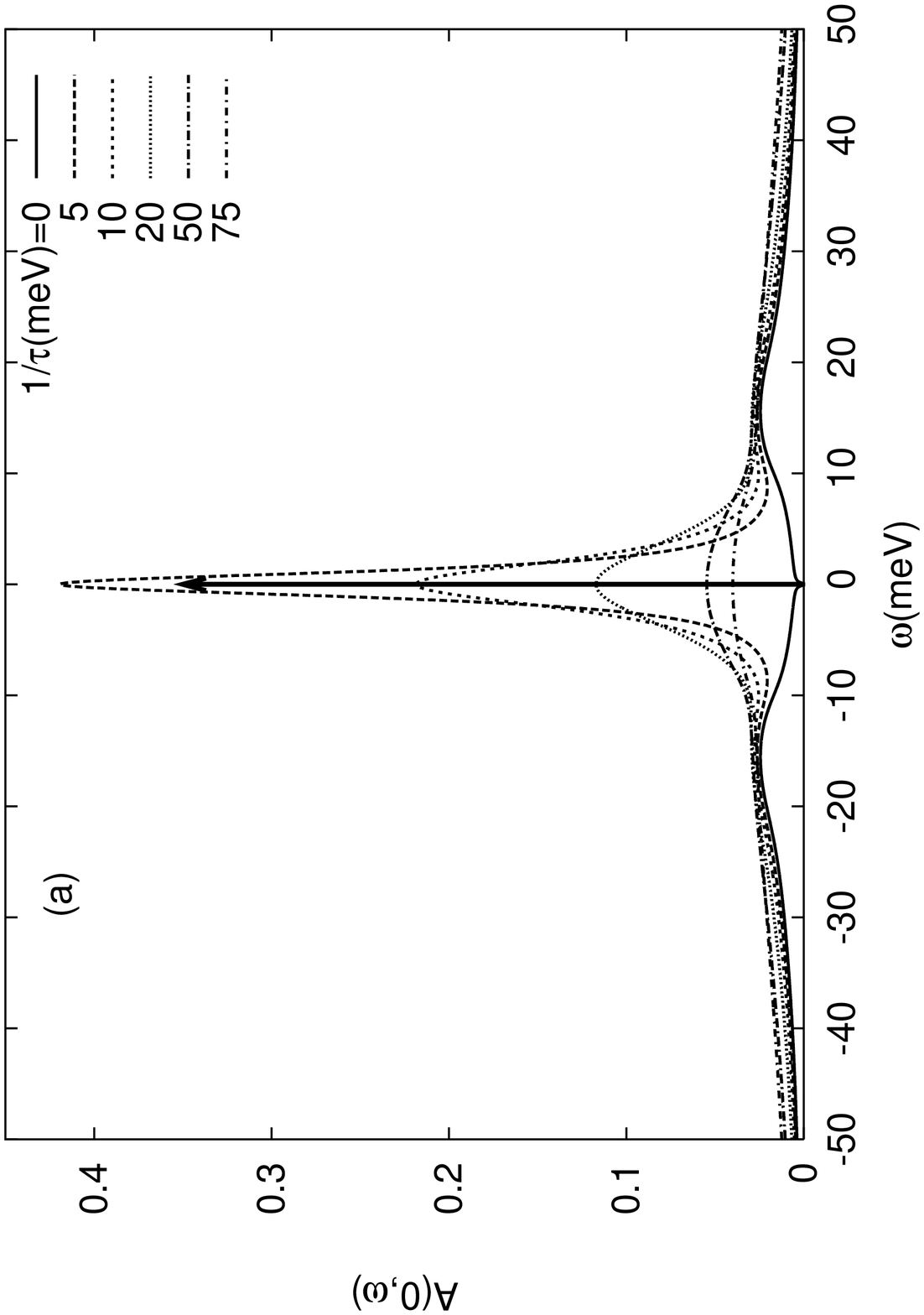,height=3.5in} \epsfig{figure=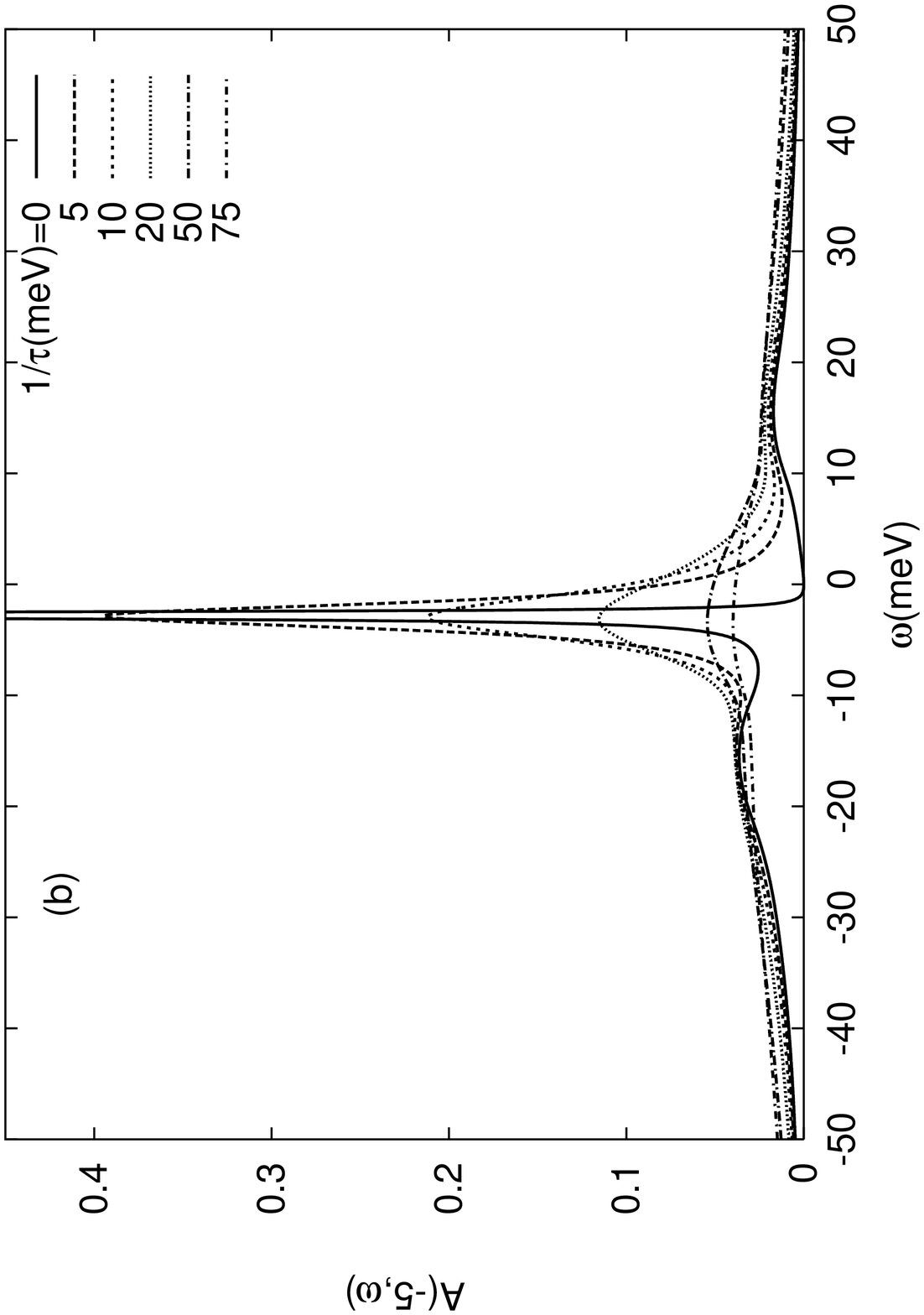,height=3.5in}
\end{turn}
\caption{The spectral function at (a) the Fermi level and (b) -5 meV for the same parameters as in Fig. 6. Note that a delta
function occurs only at the Fermi level (indicated by the arrow in (a)) when no impurity scattering is used. Increasing the
impurity scattering strength immediately broadens the delta function into a Lorentzian-like piece; the incoherent part is already
broad, and changes only quantitatively.}
\end{center}
\end{figure}
Examining the entire spectral function and/or EDOS may be a more
promising way to proceed. Of course some experimental challenges
concerning background and resolution would have to be overcome in
the meantime. In Fig.~8 we show the spectral function for (a)
$\epsilon_k = 0$ and (b) $\epsilon_k = -5$ meV. Clearly once
impurity scattering is included the separation of the quasiparticle
contribution (normally a delta-function without impurities) from the
incoherent part (normally with weight $\lambda/(1 + \lambda)$ when
the Fermi energy is infinite) will become very difficult. This is
especially true for a broad electron-phonon spectrum, particularly
one that goes to zero at zero frequency (the one used here is cut
off before zero frequency is achieved, at $\omega=0.05$meV).

\begin{figure}[tp]
\begin{center}
\begin{turn}{-90}
\epsfig{figure=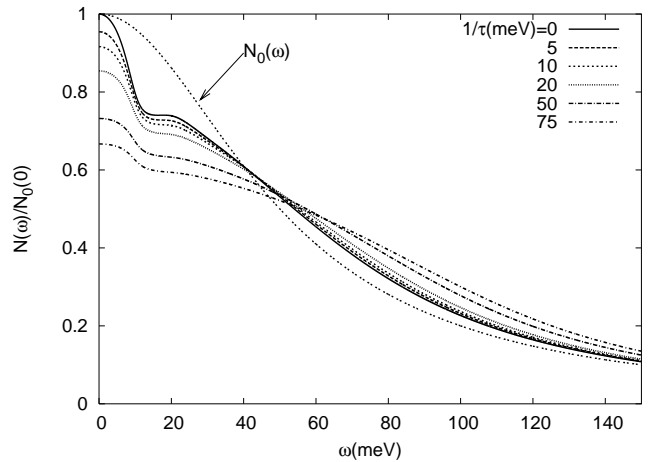,height=3.5in}
\end{turn}
\caption{The renormalized EDOS as a function of frequency for the parameters of Fig.~6. The bare EDOS is also shown. Note that
the structure near the phonon frequencies is smoothed considerably by the impurity scattering. The latter also accounts for a
considerable shift of spectral weight to higher frequency.}
\end{center}
\end{figure}
In Fig.~9, we show the renormalized EDOS, for various impurity
scattering strengths. Even if the underlying electron band structure
were known with absolute certainty it would be very difficult to
disentangle electron-phonon effects from electron-impurity effects.

All of the results presented so far utilized a Lorentzian-shaped bare
EDOS to represent the ``bare" electron spectrum. The results in
Ref.~(\onlinecite{cappelluti03}) were for a constant bare EDOS;
these can be compared with one another to see what the effect of the
bare EDOS is. Here we show a specific example in Fig.~10, for
$\lambda = 1$ and $1/\tau = 10$ meV. The low energy part tends to be
very similar, while the higher energy features experience
quantitative changes. This implies that low energy data will be
somewhat insensitive to the bare band structure, whereas high energy
data will depend quantitatively on the specific form of the band
structure. These effects will make the analysis of ARPES data
much harder as different effects are highly entangled. In addition,
the high energy part of the various single particle properties
discussed here is very much dependent on the underlying EDOS.
\begin{figure}[tp]
\begin{center}
\begin{turn}{-90}
\epsfig{figure=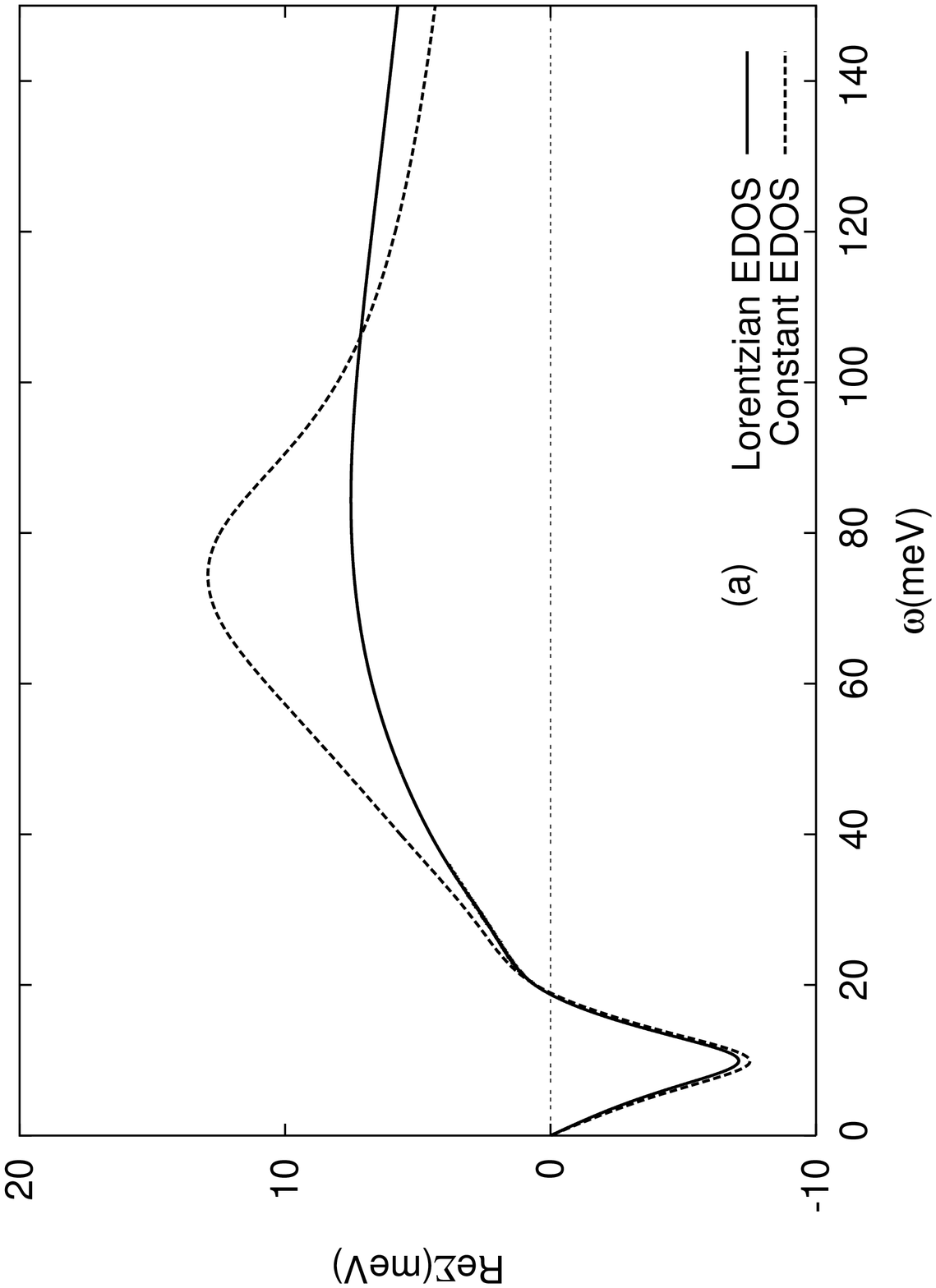,height=3.5in} \epsfig{figure=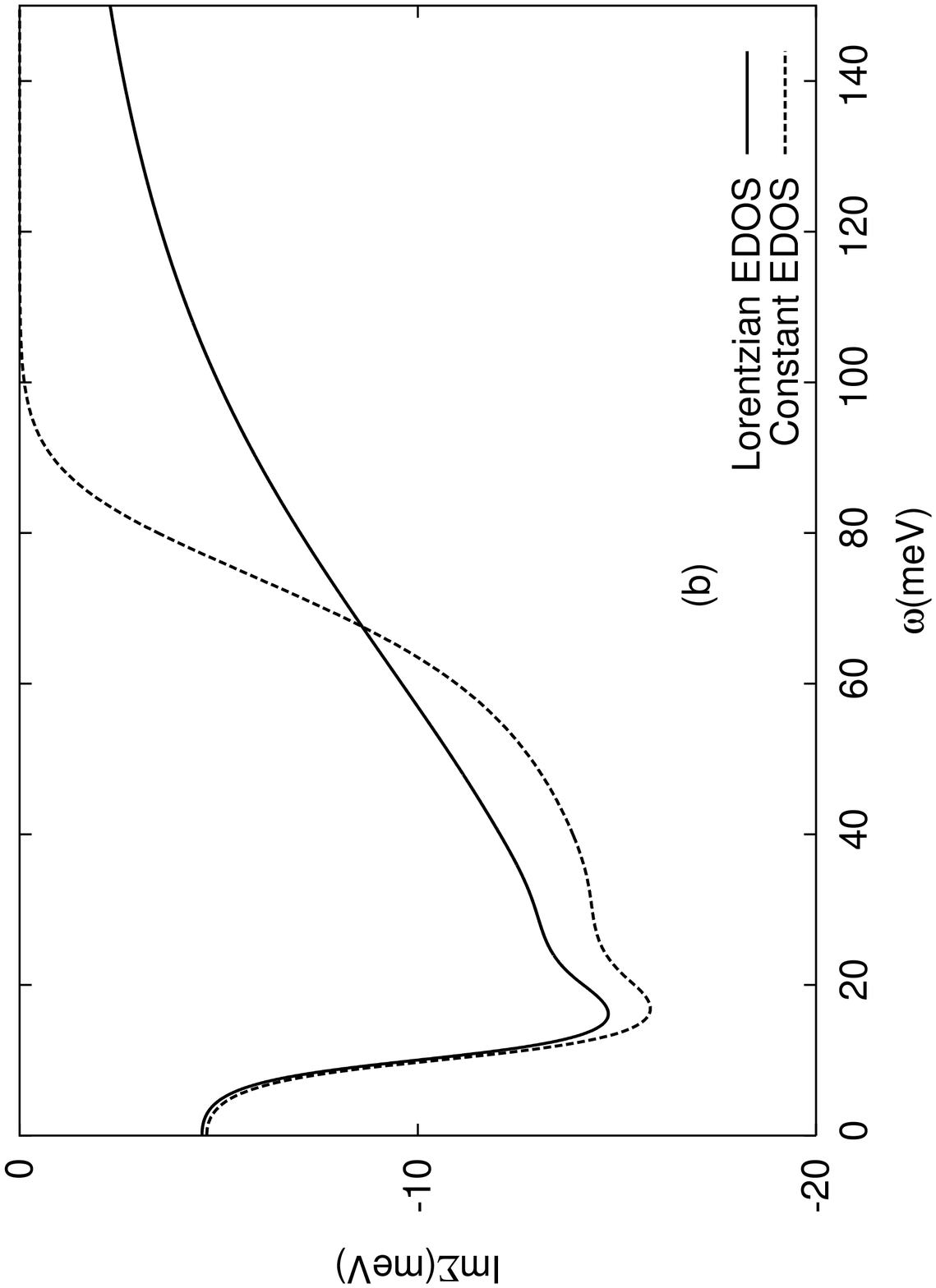,height=3.5in} \epsfig{figure=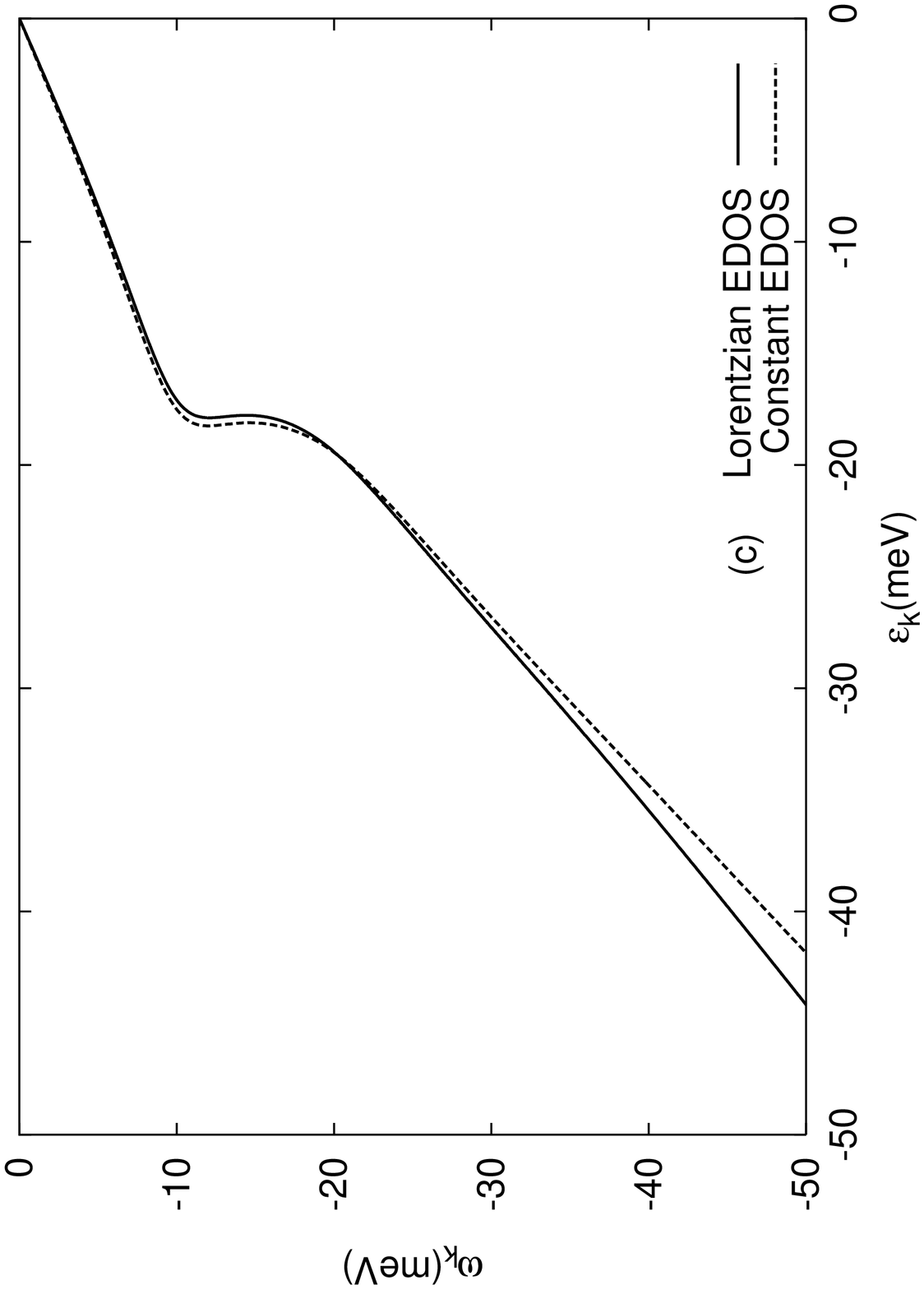,height=3.5in}
\end{turn}
\caption{A comparison of the (a) real, and (b) imaginary part of the self energy, and (c) the dispersion, for a Lorentzian (solid
curve) vs. constant (dashed curve) bare EDOS. Mostly high frequency parts change quantitatively. This means that a full inversion
of such curves to determine the underlying microscopic interactions would in practice be very difficult.}
\end{center}
\end{figure}

\section{V. SUMMARY}
\label{sec:summary} In summary, we have examined the effect of both
electron-phonon and single impurity scattering on the electron self
energy and Green function under special circumstances. These include
the presence of a finite bandwidth and phonons which are not
infinitely sharp in energy. The broadened nature of the phonon
spectrum can be due to intrinsic effects on a single phonon
(anharmonicity, interactions, etc.) that give rise to a broadened
lineshape; on the other hand they may arise from simply accounting
for the dispersion displayed by phonons in real materials. We found
that the ``smearing" effect expected from impurity scattering is
much more pronounced when such a realistic phonon spectrum is used.
Otherwise many single electron properties retain sharp features in
frequency space when an Einstein spectrum is used. Unfortunately,
this combination of finite bandwidth (usually with unknown band
structure) and uncertain degree of impurity scattering hinders the
ability to infer details concerning the electron-phonon interaction.

\begin{acknowledgments}
We acknowledge discussions with Anton Knigavko and Jules Carbotte.
This work was supported in part by the Natural Sciences and Engineering
Research Council of Canada (NSERC), by ICORE (Alberta), and by the
Canadian Institute for Advanced Research (CIAR).
\end{acknowledgments}

\bibliographystyle{prl}

\end{document}